%% file: ASM_Pure.tex
\documentclass{article}
\usepackage{amsmath,amsfonts,amssymb,amsthm}
\usepackage{tikz,forest}
\usepackage{standalone}
\DeclareMathOperator{\Ima}{Im}
\newtheorem{definition}{Definition}
\newtheorem{theorem}{Theorem}
\newtheorem{proposition}{Proposition}

\newcommand{\Z}{\mathbb{Z}}
\title{Automatic Supermartingales Acting on Sequences}
\author{Birzhan Moldagaliyev}
\begin{document}
\maketitle
\begin{abstract}
This paper describes a construction of supermartingales realized as automatic functions. A capital of supermartingales is represented using automatic capital groups~(ACG). Properties of these automatic supermartingales are then studied. Automatic supermartingales induce a notion of random infinite binary sequence. We show that the class of random sequences coincide with that of disjunctive sequences. 
\end{abstract}
\section{Introduction}
One of the earliest investigations of randomness within automata-theoretic framework is the work of Schnorr and Stimm \cite{Schnorr1972}. In that paper it is shown that automatic martingales could detect any infinite binary sequences which are not normal as defined by Borel \cite{Borel1909}, where normality refers to equidistribution of all subwords. Since then there has been numerous other studies in nature of randomness using automata-theoretic framework. In particular, Staiger \cite{Staiger1976, Staiger1998, Staiger2012} proposed an idea of regular nullsets, $\omega$-languages of Lebesgue measure $0$. Within this framework random sequences correspond to disjunctive ones as defined in \cite{Calude1997}, where disjunctivity refers to presence of all words as subwords. There is a series of works \cite{OConnor1988, Dai2004, Shen2017} where authors have introduced various automatic predicting machines, compressing machines and gambling machines. In all of these framework these machines were able to detect sequences which are not normal. One could also cite the work by Tadaki \cite{Tadaki2014} where predictability in the sense of Aulgorithmic Information Theory is introduced. There nonpredictability coincides with disjunctivity. So there is a clear dichotomy between resulting randomness classes of infinite binary sequences with regards to automatat theoretic approaches between normality and disjunctivity, the former being much more stronger condition. In this paper, we investigate the very first model proposed by Schnorr and Stimm, and try to understand its properties that providing a strength to detect sequences which are not normal. We identify this property as \emph{sequential} or \emph{local} automaticity. We then try to replace local automaticity with a global one. For this we define a notion of \emph{automatic capital group}~(ACG) which serves as a capital for martingales. We then show that under these global automaticity condition, randomness coincide with disjunctivity. \\
The rest of the paper is organized as follows. Section 2 provides necessary background for the rest of the paper and Section 3 reviews automatic martingales defined by Schnorr and Stimm. Then Section 4 introduces a notion of automatic capital groups and Section 5 introduces a notion of automatic supermartingales. In Section 6 we show that randomness in new settings correspond to disjunctivity and Section 7 offers discussions. 
\section{Background}
\subsection{Randomness via martingales}
Schnorr \cite{Schnorr1971} has suggested algorithmic martingales as a way to measure randomness of one-way infinite sequences, or $\omega$-words. One may view martingales as betting strategies. If there is algorithmic betting strategy which succeeds on a given sequence, then the sequence is assumed to showcase definite patterns. On the other hand, if no algorithmic martingale succeeds on a given sequence, then the sequence is said to be random. More formally, a martingale is function $d:\Sigma^*\to \mathbb{R}^{\ge 0}$, where $\Sigma = \{0,1\}$ a binary alphabet, having some algorithmic structure satisfying a fairness condition
\begin{equation}
\frac{1}{2}(d(x0) + d(x1)) = d(x)
\end{equation}
for all $x\in\Sigma^*$. Sometimes martingales are replaced with supermartingales, where above equality condition is replaced with inequality condition
\begin{equation}
\frac{1}{2}(d(x0) + d(x1)) \le d(x)
\end{equation}
for all $x\in\Sigma^*$. Clearly any martingale is a supermartingale, hence supermartingales are a bit more general than martingales. Moreover, they are easier to handle, due to the fact that generally inequality conditions are lass delicate when compared with equality conditions. One could then define a success of a (super)martingale on an infinite sequence $X\in\Sigma^{\mathbb{N}}$. Below $X[n]$ denotes a prefix of $X$ of length $n$. We say that (super)martingale $d$ \emph{succeeds} on $X$ if
\begin{equation}
\limsup_n (X[n]) = \infty
\end{equation}
Given a (super)martingale $d$, a collection of sequences on which $d$ succeeds is called \emph{covering region} of $d$ and denoted as $S^{\infty}[d]$. Observe that given a supermartingale $d$ with covering region $S^{\infty}[d]$, one could construct a martingale $d'$ such that $S^{\infty}[d]\subseteq S^{\infty}[d']$. This is done by adding balancing capital to children of $x0$ and $x1$ for all $x\in \Sigma^*$. This fact implies that the following definition of randomness does not depend on a distinction between martingales and supermartingales. A sequence $X$ is called \emph{random} if there is no (super)martingale succeeding on it, i.e. there is no $d$ such that $X\in S^{\infty}[d]$. 
\subsection{Automatic structures}
Assuming familiarity with regular langauges, we review a notion of automatic relation. Automatic relations~\cite{Khoussainov1995} extend a notion of regularity from languages over simpler spaces to languages over product spaces. Suppose we are given a $k$-ary relation $R\subseteq (\Sigma^*)^k$, where $\Sigma = \{0,1\}$. One might ask if there is some automatic way of computing given relation. A notion of automatic relation attempts to do that. For that, we write given $k$-tuple $t=(t_1,t_2,\ldots,t_k)\in (\Sigma^*)^k$ in a block form, also known as a convolution of $t_1,\ldots,t_k$
\begin{equation}
conv(t_1,\ldots,t_k)=
\begin{bmatrix}
t_1\\
t_2\\
\cdots\\
t_k
\end{bmatrix}
\end{equation}
To make rows homogenous, shorter rows are filled with a special symbol, say $\#$. To process such blocks, one uses finite automata which read one symbol across all rows at a time. A given relation $R$ is said to be \emph{automatic} if there is a finite automaton $M$ recognizing it. Observe that regular languages then coincide with unary automatic relations. 
\paragraph{Automatic functions} Given a notion of automatic relation, it is straightforward to define a notion of \emph{automatic function}. A function $f:(\Sigma^*)^m \to (\Sigma^*)^n$ is called automatic if its graph forms an automatic relation, i.e. $graph(f) = \{(x,f(x))\mid x\in (\Sigma^*)^m \}$ is automatic.
\paragraph{Closure under first order definition} One of the most useful tools in working with automatic relations is their closure under the first order definition~\cite{Khoussainov1995}. This property can be states as follows:
\begin{proposition}[\cite{Khoussainov1995}]
Let $R$ be a first-order definable relation from given functions $(f_1,f_2,\ldots,f_n)$ and relations $(R_1,R_2,\ldots,R_m)$. If each of these functions and relations is automatic, then $R$ is also automatic. 
\end{proposition}
\paragraph{Automaticity of mathematical structures} A notion of structure plays an important role in mathematics. A mathematical structure can be viewed as a set with certain relations defined on it. Automata theory can be used to study some of mathematical structures. A structure $\mathcal{M} = (M,f_1,f_2,\ldots,f_n,R_1,R_2,\ldots,R_m)$ where $M$ is an underlying set, $f_i$'s are functions and $R_j$'s are relations is called automatic if
\begin{itemize}
	\item $M$ is a regular language;
	\item Each $f_i$ is an automatic function;
	\item Each $R_j$ is an automatic relation.
\end{itemize}
Moreover, structures which are isomorphic to automatic structures are also called automatic. 
\subsection{FA presented group}
One of the most studied automatic structures are groups, which are referred as FA-presented groups as opposed to automatic groups in order to avoid confusion with a notion of automatic groups as in \cite{Epstein1992}. For a group $G$ to be FA-presented, it should have a presentation $(D,\circ)$ where $D$ is a regular language and group operation, $\circ$ is automatic function. Due to closure of automatic relations under first-order definition, we have that inverse function is also automatic. One of the most well-known examples of FA-presented groups is $\mathbb{Z}$. On the other hand, it is known that $\mathbb{Q}$ is not FA-presented group under either addition or multiplication \cite{Tsankov2011, Khoussainov2004}. This shows that automaticity is somehow restrictive requirement, which is cost one has to pay for nice algorithmic properties they showcase. As for notations, we are going to use additive notation for groups, with $0$ to denote an identity element. 
\section{Review of automatic martingales by Schnorr and Stimm}
Schnor and Stimm \cite{Schnorr1972} introduced a notion of automatic martingale, which was later slightly modified in \cite{Dai2004}. The basic idea behind finite state gamblers (FSG), in modified terminology can be described as follows. An FSG is a 5-tuple $M = (Q, \delta, \beta, q_0, c_0)$ where
\begin{itemize}
	\item $Q$ is a finite set of states;
	\item $\delta:Q\times \Sigma \to Q$ is a transition function;
	\item $\beta: Q\to [0,1]$ is a mapping from states to betting values.
	\item $q_0\in Q$ is the intial state;
	\item $c_0$ is the initial capital value.
\end{itemize}
At first $d(\varepsilon) = c_0$, with other values computed recursively
\begin{equation}
d(wa) = 
\begin{cases}
2\beta(\delta(q_0,w))d(w) &\text{ if }a=1 \\
2(1-\beta(\delta(q_0,w)))d(w) & \text{ otherwise }
\end{cases}
\end{equation}
Observe that computation of $d(wa)$ requires $d(w)$ along with the original string $wa$. In other words computations are performed sequentially, or locally. We term this type of automaticity \emph{sequential} or \emph{local}. In contrast, \emph{global} automaticity requires that only input for a martingale $d$ is an original string and output is corresponding capital value. Before proceeding to globally automatic martingales let us show how automatic martingales by Schnorr and Stimm formalise in the sense of automatic structures. We only work in the case when betting values are given as rational numbers. Indeed, if $\beta(Q)\subseteq \mathbb{Q}$, then there is an automatic function $d_A$, which inputs $d(w)$ along with $wa$ and outputs $d(wa)$. Let us briefly sketch how such an automatic function can be constructed.
\paragraph{Construction of $d$}
Let $Q = \{q_1,q_2,\ldots,q_n\}$ and $\beta(q_k) = \frac{a_k}{b_k}$ where $0\le a_k \le b_k$. Let $m$ be the lowest common denominator of $\{\frac{a_k}{b_k}\}_{k=1}^n$. Let $R_m = \{\frac{a}{m^l}\mid a,l\in \mathbb{Z}\}$ be an additive group of $m$-adic rationals. This group is FA-presented as shown in \cite{Nies2007}. Furthermore, multiplication by any fixed constant of the form $\frac{a}{m^l}$, where $a\in\mathbb{Z}$, is automatic for elements of $R_m$. This is due to fact that mulplication by $m,m^{-1}$ correspond to shifts in the presentation of $m$-adic rationals, which can be computed by automata. Moreover, using closure under first-order definition, one could show multiplciaiton by any fixed constant is automatic. This shows that multiplication by $2(\frac{a_k}{b_k})$ and $2(1-\frac{a_k}{b_k})$ is automatic for any $k\le n$. By combining original automaton with multiplication functions one infers that $d_A:\Sigma^*\times R_m\to R_m$ is indeed an automatic function. To ger rid of additional input of current capital value, one could represent value of $d$ as a product of betting values such as
\begin{equation}
d(wa) = c_0\prod_{i=1}^{\vert w \vert}c_k 
\end{equation}
where $c_k = 2\beta(\delta(q_0,w[k]))$ or $c_k = 2(1-\beta(\delta(q_0,w[k]))$ depending on the value of $w(k+1)$. Observe that this representation of capital is highly non-automatic. Verification of equality in values between two representations requires significant computaitonal efforts. Moreover, addition of two elements in this representation if highly nontrivial, to say the least. Ideally, given a supposed martingale $d$, it is desirable that verification that $d$ is indeed a martingale would be simple. In case of automatic $d$ with FA presented group as a capital, verification of the fact that $d$ is indeed a martingale can be done on a finite automaton due to closure of automatic relations under first-order definitions. 
	
\section{Automatic Capital Groups}
Let $C$ be FA-presented abelian group. Consider a group homomorphism $\pi:C\to \Z$ satisfying following conditions:
\begin{itemize}
	\item $\Ima \pi$ is unbounded;
	\item $\ker \pi$ is an automatic relation;
	\item $C^+ = \pi^{-1}(\Z^+)$ is an automatic relation.
\end{itemize}
A pair $(C,\pi)$ is called \emph{presentation of automatic capital group}. A map $\pi$ is called \emph{capital valuation} or \emph{valuation} for short. A group $C$ is called \emph{automatic capital group} (ACG). When we work with ACG $C$, we usually assume some valuation which is going to be clear from the context.

\subsection{Obsevations about ACG}
\paragraph{Injectivity} Let us observe that $\ker \pi$ inabove definition could be taken to be trivial. Given any ACG presentation $(C,\pi)$ we could quotient out $\ker \pi$ resulting in another FA-presented group $C'$. To be more precise, we first define an equivalence relation $\sim$ on $C$ such that:
\begin{equation}
x\sim y \Leftrightarrow x-y \in \ker \pi
\end{equation}
For each equivalence class $[x]$ we choose its length-lexicographically smallest element $x'$ as its representative. Let $C'$ be collection of representatives. Then $C'$ forms an FA-presented group because its domain can be given by an automatic relation:
\begin{equation}
x\in C' \Leftrightarrow \forall y\,(x\sim y \Rightarrow x\le_{ll}y)
\end{equation}
Furthermore, group operation is also automatic simply because, given two representatives $x',y'$ finding repsepresentative for the class $[x'y']$ can be done in automatic fashion.
\begin{equation}
x' +_{C'}y' = z' \Leftrightarrow z' = \min_{ll}\{z\mid z\sim x'y'\}
\end{equation}
The group $C'$ is isomorphic to the image $\Ima \pi$, because:
\begin{equation}
C' \simeq C/\ker \pi \simeq \Ima \pi
\end{equation}
Let us consider a restriction of $\pi$ on $C'$: $\pi' = f|_{C'}$. Then a pair $(C',\pi')$ forms an ACG presentation satisfiyng the condition that $\ker \pi'$ is trivial. Thus we can assume that a homomorphism $\pi$ is injective. From now on, we assume that $\pi$ is injective.
\paragraph{Order} Given an ACG presentation $(C,\pi)$ a group $C$ can be made into an ordered group. Let us define a following natural ordering:
\begin{equation}
x > y \Leftrightarrow x - y \in C^+
\end{equation}
Clearly this order is automatic, due to automaticity of $C^+$. Having an ordering, we can talk about upper bounds and least upper bounds. Given a subset $D\subseteq C$, we can look at set of its upper bounds:
\begin{equation}
U(D) = \{x\in C\mid \forall d\in D\,(x\ge d)\}
\end{equation}
If $U(D)$ is empty, we say that $D$ is not upper bounded and write $\limsup(D)=\infty$. On the other hand, if $U(D)=C$, then $D$ must be an empty set, and write $
\limsup (D) = -\infty$. In the remaining case the order type of $D$ must be equal to that of natural numbers. Hence, $U(D)$ has the least element. Hence we define
\begin{equation}
\limsup(D) = \min(U(D))
\end{equation}
Due to its first order definition, $\limsup$ has a nice behavior in terms of automaticity. The following proposition demonstrates some of the conveniences of working with $\limsup$ in this context.
\begin{proposition}
Let $R\subseteq C\times C$ be an automatic relation. Consider $Im_R(c) = \{d\mid R(c,d)\}$ image of an element $c$ under $R$. Let us define a function $f:C\to C\cup\{\pm \infty\}$ as
\[
f(c) = \limsup (Im_R(c))
\]
Then $f$ is an automatic function.
\end{proposition}
\begin{proof}
Given an element $c\in C$ we need to decide if $Im_R(c)$ is nonempty first. It can be done in the first-order fashion: $\exists d(R(c,d))$. In case $Im_R(c)$ happens to be empty, we set $f(c) = -\infty$. Then we need to check if $Im_R(c)$ is bounded. Again this can be checked in the first-order fashion by writing
\begin{equation}
\limsup(Im_R(c))<\infty \Leftrightarrow \exists d'\, \forall d\,(R(c,d)\Rightarrow d\le d')
\end{equation}
If $Im_R(c)$ happens to be unbounded, we write $f(c) = \infty$. Finally we set $f(c) = \min U(Im_R(c))$ in the remaining case. Observe that last expression is also first-order definable. Since all of above conditions involve first-order formulas in terms of $c$, this means that computation of $f$ is automatic uniformly in $c$. 
\end{proof}
Let us try to clarify the reason behind considering ACG presentations. Classical martingales usually work in ordered fields like rational numbers $\mathbb{Q}$ or dyadic rationals $\mathbb{Q}_2$. Usage of $\mathbb{Q}$ as capital values is hardly feasible, due to result in \cite{Tsankov2011} that $(\mathbb{Q},+)$ does not form an FA-presented group. On the other hand standard representations of dyadic rationals or even integers in binary are found to be too rigid to work with. For this reason we wanted to find an appropriate FA-presented groups which have features of order and boundedness to define a notion of automatic (super)martingales.

\section{Automatic Supermartingales}
Let $(C,\pi)$ be an ACG presentation. Let us define a notion of \emph{automatic supermartingale} (ASM).
\begin{definition}
A mapping $d:\{0,1\}^*\to C$ is called automatic supermartingale (ASM) if it satisfies following conditions
\begin{enumerate}
	\item $d$ is an automatic function;
	\item $d(\sigma 0) + d(\sigma 1) \le 2d(\sigma) = d(\sigma) + d(\sigma)$;
	\item $d(\sigma)\ge 0_C, \quad \forall \sigma\in\{0,1\}^*$.
\end{enumerate}
\end{definition}
The second condition is a type of fairness condition used for supermartingales. A third condition is an adoption of the classical condition that martingale values should be nonnegative. One could define automatic martingale (AM) be replacing inequality in the second condition by equality. However, we prefer to work with ASM's due to ease of working with them. We can define a success criteria in usual way. 
\paragraph{Success criteria} Let $X:\mathbb{N}\to \{0,1\}$ be an infinite binary sequence. Given $X = x_1x_2x_3\ldots$, $X[k] = x_1x_2\ldots x_k$ denotes prefix of $x$ of length $k$. We say that ASM $d$ succeeds on $X$ if
\begin{equation}
\limsup_k\{d(X[k]) \} = \infty
\end{equation}
Finally we are ready to define a notion of random sequences.
\paragraph{Random sequence} We say that a sequence $X$ is ASM random if there is no ASM succeeding on $X$. Observe that in this definition, we do not fix class of ACG groups used by automatic supermartingales. They are allowed to use any kind of ACG presentations. This fact is going to play a crucial role later. 

\subsection{Observations on ASMs}
Let us make few observations about ASMs. In theory of algorithmic randomness, it is common to add two (super)martingales in order to obtain a new (super)martingale which might have nice properties. One could ask if it is possible to add two ASMs. Let us consider two ASMs $d_1:\Sigma^*\to C_1$ and $d_2:\Sigma^*\to C_2$ with capital presentations $(C_1,\pi_1)$ and $(C_2,\pi_2)$ respectively. Since $C_1$ and $C_2$ might be different groups it is not clear if addition of $d_1$ and $d_2$ makes any sense. One possibility to approach this problem is by transferring everything into $\Z$ via valuation maps. In other words, we say that $d:\Sigma^*\to C$ with capital presentation $C,\pi$ is an addition of $d_1$ and $d_2$ if
\begin{equation}
\pi\circ d(\sigma) = \pi_1\circ d_1(\sigma) + \pi_2 d_2(\sigma), \quad \forall \sigma\in \Sigma^*
\end{equation}
One could immediately see that such $d$ might not be unique due to different possible presentation. One possibility to address this issue to define an equivalence of ASMs by letting $d_1\sim d_2$ if 
\begin{equation}
\forall \sigma\in \Sigma^*(\pi_1\circ d_1(\sigma) = \pi_2\circ d_2(\sigma))
\end{equation}
Then we have that addition of two ASMs is unique up to equivalence. However, uniqueness is of small concern for us, while existence seem to be more daunting challenge. One might ask when does such $d=d_1+d_2$ exist. We could present a sufficient conditions for existence of such $d$. We need few definitions to get started. 
\begin{definition}[Compatibility of capital presentation]
Two capital presentations $(C_1,\pi_1)$ and $(C_2,\pi_2)$ are called compatible if following relations are automatic
\begin{enumerate}
	\item $R_= = \{(x,y)\in C_1\times C_2\mid \pi_1(x) = \pi_2(y) \}$;
	\item $R_< = \{(x,y)\in C_1\times C_2\mid \pi_1(x)<\pi_2(y) \}$.
\end{enumerate}
\end{definition}
Now we are ready to state the sufficient condition.
\begin{proposition}
Let $d_1:\Sigma^*\to C_1$ and $d_2:\Sigma^*\to C_2$ be two ASMs with corresponding capital presentations $(C_1,\pi_1)$ and $(C_2,\pi_2)$. Suppose that $(C_1,\pi_1)$ and $(C_2,\pi_2)$ are compatible presentations. Then $d=d_1+d_2$ exists.
\end{proposition}
\begin{proof}
We start by constructing an ACG presentation $(C,\pi)$ for $d$. Let $C = C_1\times C_2$ be a product of two groups $C_1$ and $C_2$. Since both of them are FA-presented, $C$ is also FA-presented \cite{Blumensath2000}. Consider a following valuation map $\pi:C\to \Z$
\begin{equation}
\pi(x,y) = \pi_1(x) + \pi_2(y)
\end{equation}
where $x\in C_1$ and $y\in C_2$. We can easily verify that $(C,\pi)$ forms an ACG presentation. Image of $C$ is unbounded due ot the fact that $\pi_1(C_1)\subseteq \pi(C)$ following from observation that $\pi(x,0_{C_2}) = \pi_1(x)$. Kernel of $\pi$ is an automatic relation because
\begin{equation}
(x,y)\in \ker \pi \Leftrightarrow \pi_1(x)+\pi_2(y) = 0 \Leftrightarrow (x,-y)\in R_=
\end{equation}
Furthermore $C^+ = \pi^{-1}(\Z^+)$ is an automatic relation because
\begin{equation}
(x,y)\in C^+ \Leftrightarrow \pi_1(x) + \pi_2(y)>0 \Leftrightarrow (-x,y)\in R_<
\end{equation}
We have finally shown that $(C,\pi)$ is an ACG presentation. Now we need to construct ASM itself. Let us define $d:\Sigma^*\to C$ as follows
\begin{equation}
d(\sigma) = (d_1(\sigma),d_2(\sigma)),\quad \forall \sigma\in \Sigma^*
\end{equation}
Now we need to verify that $d$ is indeed an ASM. Since $d_1$ and $d_2$ are automatic functions, $d$ is also automatic, being the convolution of former two. Since $d_1$ and $d_2$ satisfy fairness conditions for supermartingales, so does $d$. More explicity
\begin{align}
d(\sigma 0) + d(\sigma 1) &= (d_1(\sigma 0), d_2(\sigma 0)) + (d_1(\sigma 1),d_2(\sigma 1))\\        
                                           &= (d_1(\sigma 0)+d_1(\sigma 1), d_2(\sigma_0) + d_2(\sigma_1))\\
                                           &\le (2d_1(\sigma), 2d_2(\sigma))\\
                                           &= 2d(\sigma), \quad \forall \sigma\in \Sigma^*
\end{align}
As for nonnegativity, we have that
\begin{align}
d(\sigma) &=(d_1(\sigma),d_2(\sigma))\\
                &\ge (0_{C_1},0_{C_2})\\
                &= 0_C
\end{align}
We have finally shown that $d$ is ASM with ACG presentation $(C,\pi)$. By the construction we conclude that the condition for being addition of $d_1$ and $d_2$ holds for $d$
\begin{equation}
\pi\circ d(\sigma) = \pi_1\circ d_1(\sigma) + \pi_2 \circ d_2(\sigma)
\end{equation}
Hence an addition of $d_1$ and $d_2$ exists in the form of $d$. 
\end{proof}

\section{Characterization of random sequences}
This section characterizes randomness of infinite binary sequences given by ASM. For that we need to recall the definition of disjunctivity \cite{Calude1997}.
\begin{definition}[Disjunctive sequence \cite{Calude1997}]
An infinite sequence $X$ is said to be disjunctive if any word $w\in \Sigma^*$ appears in $X$ as a subword. 
\end{definition}
Now we can state the characterization of ASM random sequences.
\begin{theorem}
A sequence $X$ is ASM random if and only if it is disjunctive
\end{theorem}
\begin{proof}
As it is easier to work with negations in this case, we consider a following equivalent restatement:
\begin{center}
$X$ is not ASM random if and only if $X$ is not disjunctive.
\end{center}
Let us verify both directions of the reformulated theorem.
\paragraph{Forward direction} As $X$ is not ASM random, there is an ASM $d:\Sigma^*\to C$ with capital presentation $(C,\pi)$ succeeding on $X$. Let $S^{\infty}[d] = \{Z\mid d\text{ suceeds on } Z \}$ be a collection of sequences $d$ succeeds on. We are going to show that $S^{\infty}[d]$ is deterministic B\"{u}chi-recognizable $\omega$-language of measure $0$. Then it would follow that $X$ is not disjunctive according to the result in \cite{Staiger1976}. Observe that $\pi\circ d:\Sigma^*\to \Z$ is an ordinary supermartingale and $S^{\infty}[\pi\circ d] = S^{\infty}[d]$. As $\pi\circ d$ is an ordinary martingale as in theory of algorithmic randomness, we have that  $\mu(S^{\infty}[\pi\circ d]) = 0$ as shown in \cite{AmbosSpies1997}, thus we infer that $\mu(S^{\infty}[d]) = 0$. So we are left to show that $S^{\infty}[d]$ is recognized by deterministic B\"{u}chi automata. In other words we need to show an existence of regular language $R$ such that
\begin{equation}
Z\in S^{\infty}[d] \Leftrightarrow \vert Pref(Z)\cap R \vert = \infty
\end{equation}
where $Pref(Z)$ refers to the collection of all prefixes of $Z$. Let us define $R$ as follows
\begin{equation}
R = \{\sigma\mid \forall \tau\prec \sigma(d(\tau)<d(\sigma)\}
\end{equation}
Above language is regular due to first-order definability property. Suppose $\vert Pref(Z)\cap R\vert = \infty$, then there are prefixes $\{\sigma_i \}_{i\in \mathbb{N}}$ such that $d(\sigma_{i+1})>d(\sigma_i)$ for all $i\in\mathbb{N}$. Since $\{d(\sigma_i)\}_{i\in\mathbb{N}}$ forms an ascending infinite chain in $C^{\ge 0}$, $\limsup_i\{d(\sigma_i) \}=\infty$ because order type of $C^{\ge 0}$ is the same as that of $\mathbb{N}$. It implies that $Z\in S^{\infty}[d]$. On the other hand, suppose that $\vert Pref(Z)\cap R\vert < \infty$. Then there is the longest prefix $\sigma\in R$. This means that for any $\tau\in Pref(X)$ extending $\sigma$, we must have $d(\tau)\le d(\sigma)$. Otherwise, we would get a contradiction with maximality of $\sigma$. 

\paragraph{Converse direction} Let $X$ be a nondisjunctive sequence so that it does not contain a string $w$ as a subword. We wish to construct an ASM $d:\Sigma^*\to C$ with captial presentation $(C,\pi)$ so that $d$ succeeds on $X$. The basic idea behind the construction is redistribution of capital from strings containing $w$ to strings that do not. Though an idea is fairly simple, ensuring automaticity of $d$ becomes the main challenge of the construction. To give a taste of an idea behind the construction, let us consider a simple example where $w = 00$. There are $4$ string of length $2$, so given a sequence $\sigma$ of length $2$, we want $d$ to output following values
\begin{equation}
d(\sigma)=
\begin{cases}
\frac{4}{3}. & \text{ if } \sigma \neq 00\\
0                 & \text{ otherwise}
\end{cases}
\end{equation}
As for strings of length $0$ and $1$, we backpropagate values of $d$ using a fairness condition for martingales: $2d(\sigma) = d(\sigma 0) + d(\sigma 1)$. So we get a following tree, where capital values for each string are written inside corresponding circles:
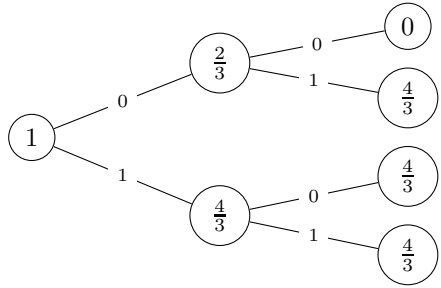
\begin{center}
\includestandalone{Tree}
\end{center}
Above procedure allows to increase initial capital by the factor of $\frac{4}{3}$ on the block of length two. By iterating this procedure three times, the capital increases by a factor of $\frac{64}{27}>2$ on the block of length $6$. Multiplicating above capital values by $27$, capital values become integer-valued. By replacing value of empty string, $\varepsilon$, with $32$ instead of $27$ we arrive at the supermartingale which doubles its initial value within a block of length $6$. By replicating the last supermartingale several times, a desired martingale succeeding on sequences not containing $00$ is obtained.
\paragraph{Generalization of the above idea} Suppose that a sequence $X$ does not contain a word $w$ on length $n$. Let $q = \frac{2^n}{2^n-1}>1$ and $d:\Sigma^n\to \mathbb{Q}$ be a function such that $d(\sigma) = 1_{\{\sigma\neq w\}}q$, here $1_{A}$ denotes an indicator function for the event $A$. By backpropagating values using a fairness condition $2d(\sigma) = d(\sigma 0) + d(\sigma 1)$, we obtain a partial martingale with a norm $1$. Let $k$ be the smallest integer such that $q^k \ge 2$. By replicating above procedure $k$ times, one obtains a partial martingale such that given a word $\sigma = \sigma_1\ldots \sigma_k$ of $k$ blocks on length $n$, we have
\begin{equation}
d(\sigma_1\ldots \sigma_k) = 1_{\{\sigma_1\neq w\} }\ldots 1_{\{\sigma_k\neq w\} } q^k
\end{equation}
By multiplying capital values obtained so far by $(2^n-1)^k$ martingale becomes integer valued. Furthermore, replacing capital value of empty string with greater value of $2^{nk-1}$ one obtains a supermartingale which doubles its initial value within the block of length $nk$. Let us denote this partial supermartingale $d$. The next step is extension of $d$ into a full supermartingale. Given any word $\tau$, it can be written $\tau = \sigma_0\ldots \sigma_s$ such that $\vert \sigma_i \vert = nk$ for $i<s$ and $0\le \vert \sigma_s \vert < nk$. The value of $d$ on $\sigma$ is then defined as
\begin{equation}
d(\tau) = 
\begin{cases}
2^s d(\sigma_s), &\text{ if } d(\sigma_i)\neq 0, \text{ for all }i<s\\
0 & \text{ otherwise}
\end{cases} 
\end{equation}
Let us verify that $d$ is indeed a supermartingale. Since $d$ forms a partial supermartingale on words of length up to $nk$, construction ensures the supermartingale condition of $2d(\sigma)\ge d(\sigma 0) + d(\sigma 1)$ is preserved for $\tau$ with $\vert \sigma_s \vert \neq nk-1$. So we have to consider the case $\vert \sigma_s \vert = nk-1$ only. Observe that
\begin{equation}
d(\tau b) = 
\begin{cases}
2^{s+1} d(\varepsilon), & \text{ if } d(\sigma_i)\neq 0, \, d(\sigma_s b)\neq 0\\
0 & \text{ otherwise}
\end{cases}\quad \text{ for } b\in\{0,1 \}
\end{equation}
In case $d(\sigma_i)=0$ for some $i<s$, then $d(\tau) = d(\tau0) =d(\tau 1) = 0$, so the supermartingale condition is satisfied. Assume that $d(\sigma_i)\neq 0$ for all $i<s$. Then we have
\begin{equation}
d(\tau b) = 2^s d(\sigma_s b), \text{ for } b\in \{0,1\}
\end{equation}
Since $d$ satisfies the supermartingale property on words of length $nk-1$, we have that it also satisfies the supermartingale property on the word $\tau$. 
\paragraph{Incorporating automaticity}
Observe that the discussions up to this point did not involve any notion of automaticity. In next phase of the construction we equip obtained supermartingale $d$ with automatic structure. First we build an ACG presentation $(C,\pi)$. We start with binary representation of integers
\[
m = (-1)^p(a_0 + 2a_1 + \ldots + 2^ta_t), \text{ where } a_t \neq 0
\]
Let us consider a representation of integers where $a_i$'s are placed $nk$ distance apart separated by $0$'s. In this representation $m$ looks like
\begin{equation}
m \to (a_00\ldots0)\ldots (a_t 0\ldots 0)p
\end{equation}
where $p\in \Sigma$ with $0$ denoting positive integers, while $1$ denotes negative integers. Let us call this representation $nk$-separated binary representation. Since integers under addition form FA-presented group using binary representation, they also form FA-presented group under $nk$-separated representation. Addition of two numbers in $nk$-separated representation is performed in the same way as in binary representation, except for the fact that one ignores separating $0$'s. Let us denote a group of integers under $nk$-seprated representation $C$. Consider a canonical mapping $\pi:C\to \mathbb{Z}$
\begin{equation}
\pi(a_00^{n-1}\ldots a_t0^{n-1}p) = (-1)^p(a_0 + 2a_1 + \ldots + 2^ta_t)
\end{equation}
Since the map $\pi$ removes all redundant $0$'s, it can be thought as a contraction mapping. On the other hand $\pi'=\pi^{-1}$ is an expansion mapping. Let us show that $(C,\pi)$ forms a ACG presentation. We go through conditions one-by-one
\begin{enumerate}
	\item Clearly, $\pi(C)=\mathbb{Z}$, hence it is unbounded.
	\item $\ker \pi = 0_C = \varepsilon$ is trivial, hence regular.
	\item $C^+ = \pi^{-1}(\Z^+) = (\{0,1\}0^{nk-1})^*\cdot(10^{nk-1})\cdot 0$ which is clearly a regular language. 
\end{enumerate}
Having constructed appropriate ACG presentation, we are left with specifying appropriate ASM $d_A:\Sigma^*\to C$ which is based on the supermartingale $d$ constructed previously. Given any sequence $\tau$ we make use of early technique of dividing it into blocks of size $nk$, i.e. $\tau = \sigma_0\sigma_1\ldots\sigma_s$, where $\vert\sigma_i\vert = nk$ for $i<s$ and $0\le \vert \sigma_s\vert < nk$. We then defined $d_A:\Sigma^*\to C$ as follows
\begin{equation}
d_A(\tau) = \pi'(d(\tau)), \text{ for all } \tau\in\Sigma^*
\end{equation}
Recall that $d(\tau)$ is computed as follows
\begin{equation}
d(\tau) = 
\begin{cases}
2^s d(\sigma_s), &\text{ if } \sigma_i\neq w, \text{ for all }i<s\\
0 & \text{ otherwise}
\end{cases} 
\end{equation}
where $\tau = \sigma_0\ldots \sigma_s$. Observe that verification of  $\sigma_i\neq w$ for all $i<s$ can be performed on a finite automaton. If this is a case, then value of $d(\tau)$ is only dependent on $s$ and $d(\sigma_s)$. Viewing $\tau$ and $d_A(\tau)$ as inputs, we then have a following picture
\begin{equation}
\begin{matrix}
\sigma_1 & \sigma_2 & \ldots & \sigma_{s-1} & \sigma_s\\
0^{nk}    &  0^{nk}     & \ldots & 0^{nk}           & d_A(\sigma_s)
\end{matrix}
\end{equation}
Since there only finitely many possibilities for $\sigma_s$, above mapping can be encoded into a finite automaton. This shows that $d_A$ is an automatic function.
\end{proof}

\section{Discussions}
In this paper we have reviewed the construction of automatic martingales by Schnorr and Stimm as automatic structures. We have observed that the construction uses sequential or local automaticity. We then have introduced a notion of global (super)martingales. In order to do this, automatic capital groups were defined. At last, it is shown that under new framework randomness coincides with disjunctivity.

\end{document}

%% file: Tree.tex
\forestset{
default preamble={
	for tree = {circle,draw, grow=0, reversed}
}
}

\begin{forest}
for tree = {l = 2.5cm}
[$1$
[$\frac{2}{3}$, edge label = {node[midway,fill=white,font=\scriptsize]{0}}
[$0$, edge label = {node[midway,fill=white,font=\scriptsize]{0}}][$\frac{4}{3}$, edge label = {node[midway,fill=white,font=\scriptsize]{1}}]
]
[$\frac{4}{3}$, edge label = {node[midway,fill=white,font=\scriptsize]{1}}
[$\frac{4}{3}$, edge label = {node[midway,fill=white,font=\scriptsize]{0}}][$\frac{4}{3}$, edge label = {node[midway,fill=white,font=\scriptsize]{1}}]
]
]
\end{forest}